\documentclass[12pt,preprint]{aastex}

\newcommand{\text}[1]{\rm #1}

\slugcomment{}

\shorttitle{Numerical Analysis on Standing Accretion Shock Instability}
\shortauthors{Ohnishi et al.}

\begin{document}


\title{Numerical Analysis on Standing Accretion Shock Instability with
Neutrino Heating in the Supernova Cores}

\author{Naofumi Ohnishi\altaffilmark{1}, Kei Kotake \altaffilmark{2},
and Shoichi Yamada\altaffilmark{2,3}}
\affil{$^1$Department of Aerospace Engineering, Tohoku University,
6-6-01 Aramaki-Aza-Aoba, Aoba-ku, Sendai, 980-8579, Japan}
\email{ohnishi@cfd.mech.tohoku.ac.jp}
\affil{$^2$Science \& Engineering, Waseda University, 3-4-1 Okubo, Shinjuku,
Tokyo, 169-8555, Japan}
\email{kkotake@heap.phys.waseda.ac.jp}
\affil{$^3$Advanced Research Institute for Science and Engineering, Waseda University, 3-4-1 Okubo, Shinjuku,
Tokyo, 169-8555, Japan}
\email{shoichi@waseda.jp}

\begin{abstract}
 We have numerically studied the instability of the spherically symmetric 
standing accretion shock wave against non-spherical perturbations. We have 
in mind the application to the collapse-driven supernovae in the post bounce phase, 
where the prompt shock wave generated by core bounce is commonly stalled. We take 
an experimental stand point in this paper. Using spherically symmetric, completely
steady, shocked accretion flows as unperturbed states, we have clearly observed both 
the linear growth and the subsequent nonlinear saturation of the instability. 
In so doing, we have employed a realistic equation of state together with heating 
and cooling via neutrino reactions with nucleons. We have done a mode analysis based on  
the spherical harmonics decomposition and found that the modes with $\ell=1, 2$ are 
dominant not only in the linear regime, but also after the nonlinear couplings generate 
various modes and the saturation occurs. Varying the neutrino luminosity, we have 
constructed the unperturbed states both with and without a negative entropy-gradient. 
We have found that in both cases the growth of the instability is similar, 
suggesting the convection does not play a dominant role, which also appears to be supported by
the recent linear analysis of the convection in accretion flows by Foglizzo et al. 
The real part of the eigen frequency seems to be mainly determined by the advection time 
rather than by the sound-crossing time. Whatever the cause may be, the instability 
is favorable for the shock revival.
\end{abstract}

\clearpage

\section{Introduction}

Over the last decades, it has been observationally suggested 
that the core-collapse supernovae are generally aspherical \citep{wang96,wang01,wang02}. 
The most unequivocal example is SN1987A. Recent HST
images of SN1987A are directly showing that the expanding envelope is 
indeed elliptical with the long axis aligned with the rotation axis inferred
from the ring. The aspect ratio and the position angle of the symmetry axis are 
consistent with those predicted earlier from the observations of speckle and 
linear polarization. What is more, the linear polarization became greater as the time 
passed \citep{wang01,leonard01}, the fact which has been used to argue that 
the central engine of the explosion is responsible for the
non-sphericity \citep{wheeler,wheeler00}.
Rather common detections of the linear polarization in other core-collapse supernovae 
seem to suggest, as mentioned at the beginning, that the core-collapse supernovae 
are globally asymmetric in general.

Various physical ingredients have been proposed as a possible cause of the asymmetry
so far: convections below the shock wave as well as inside the proto neutron star
(\citet{herant_94,burrows_95,jankamueller96}, see for a review \citet{janka_01}),
density inhomogeneities produced prior to core-collapse \citep{burohey,fryerkick},
rapid rotation of the core \citep{monch,yama94,shimi01,kotake,yamasaki,walder},
and magnetic fields \citep{ard,akiyama,kotakemhd,yamasawa,kotakekick}
(see for a review, \citet{kotake_rev} and references therein).
These studies have also been motivated by the interest in the consequences of the asymmetric 
motions to the explosion mechanism itself, since it has been expected that the deviation
from the spherical symmetry will be helpful for the shock revival one way or another.

Recently, \citet{blondin_03} demonstrated numerically that yet another hydrodynamical
mechanism may be conspiring to drive non-spherical motions in the flow below the shock wave.
The so-called standing accretion shock instability (SASI) is supposed to be a non-local 
hydrodynamical instability possibly caused by the cycle of the inward advection of velocity- and 
entropy-perturbations and outward propagation of acoustic waves, with fluctuations amplified 
after each cycle. This mechanism of SASI was originally studied in linear analysis 
by \citet{foglizzo01,foglizzo02} in the context of accreting black holes (see also \citet{HC}).
Adding small non-spherical perturbations to the spherically symmetric, isentropic, steady, 
post-shock accretion flows, \citet{blondin_03} found in their numerical simulations that 
the perturbations grow up to the nonlinear regime with clear dominance of 
$\ell=1$ at first and $\ell=2$ later, leading to the global deformation of the shock wave.
Here $\ell$ stands for the azimuthal index of the Legendre polynomials. One lesson to learn
is that we should not impose the symmetry with respect to the equatorial plane in the simulations.

As mentioned already, since the large deviation from the spherical symmetry may have an 
important consequence to the explosion itself, the finding of \citet{blondin_03} has
much interest of other researchers. In their first paper \citep{blondin_03}, 
the neutrino heating and cooling are entirely ignored and the flow is assumed to be isentropic. 
In the recent paper \citep{blondin_05}, the authors took into account the cooling term of a simple 
analytic form just as in \citet{HC}, but no heating included yet. On the other hand, 
\citet{scheck_04} demonstrated that similar asymmetric motions 
with no equatorial symmetry 
occur in their most realistic numerical models. 
Although their results show that the 
neutrino processes will not nullify SASI, the growths and saturations of individual modes
under neutrino-irradiation are not clear, since they used highly complicated flows 
as an underlying model.

Our standing point in this paper is somewhere in between these works. We study SASI 
by 2D axisymmetric hydrodynamical simulations. Although we have in mind the application 
to the supernova core in the shock-stagnation phase, we take an experimental stance 
as in \citet{blondin_03}. On the one hand, we employ a realistic
equation of state \citep{shen98} 
and take into account the heating and cooling of matter via neutrino emissions and absorptions 
on nucleons. As an underlying model, on the other hand, we utilize the spherically symmetric, 
steady, shocked accretion flows \citep{yamasaki}, which is stable against radial perturbations.  
Although this is certainly a crude approximation to what we found in the realistic simulations,
it will enable us to do clear mode-analyses from the linear growths through the nonlinear 
couplings among various modes up to the eventual saturation of SASI. Due to the neutrino-heating, 
some initial models have a convectively unstable region in the classical sense 
(see \citet{foglizzo05}) and SASI is inevitably mixed with convection. By lowering 
the neutrino luminosity, however, we can also construct models with no convectively unstable 
region. By comparing these models, we can assess the relative strength of these instabilities.
We also discuss the implications that SASI might have for the shock revival. 

The plan of this paper is as follows.
We describe the numerical methods in section \ref{sec2}.
The main numerical results are shown in section \ref{sec3}.
We conclude this paper in section \ref{sec4}.

\section{Numerical Methods and Models\label{sec2}}

\subsection{Basic Equations}

Assuming the axisymmetry of the system in this paper, we numerically study the growth of the non-spherical 
instability in the accretion flow through the shock wave onto the protoneutron star. The unperturbed steady 
accretion flows and the shock waves are assumed to be spherically symmetric. We take into account the 
heating and cooling of accreting matter via neutrino absorptions and emissions by free nucleons.
Only the region outside the neutrino sphere is considered.

The basic evolution equations are written as follows, 
\begin{equation}
 \frac{d\rho}{dt} + \rho \nabla \cdot \mbox{\boldmath$v$} = 0,
\end{equation}
\begin{equation}
 \rho \frac{d \mbox{\boldmath$v$}}{dt} = - \nabla P - \rho\nabla\Phi,
\end{equation}
\begin{equation}
 \rho \frac{d}{dt}\displaystyle{\Bigl(\frac{e}{\rho}\Bigr)}
  = - P \nabla \cdot \mbox{\boldmath$v$} + Q_{\text{E}},
  \label{eq:energy}
\end{equation}
\begin{equation}
 \frac{dY_{\rm e}}{dt} = Q_{\text{N}},
  \label{eq:ye_flow}
\end{equation}
\begin{equation}
 \Phi = - \frac{G M_{\rm in}}{r},
  \label{eq:domain_g}
\end{equation}
where $\rho, \mbox{\boldmath$v$}, e, P, Y_{\rm e}, \Phi$ are
density, velocity, internal energy, pressure, electron fraction,
and gravitational potential of the central object, respectively. The self-gravity
of matter in the accretion flow is ignored. 
$Q_{\rm E}$ and $Q_{\rm N}$ are related with the interactions with neutrinos
and are explained in more detail in the next section.
We denote the Lagrangian derivative as $d/dt$ and $r$ is the radius.

The numerical code for hydrodynamic computations employed
in this paper is based on the ZEUS-2D \citep{stone},
which is an Eulerian code based on the finite-difference method
and employs an artificial viscosity of von Neumann and Richtmyer type
to capture shocks. We have made several major changes to the base code
to include the microphysics as described in the following sections.
First, we have added the equation for electron fraction
(Eq.~(\ref{eq:ye_flow})), which is solved in the operator-splitting fashion.
Second, we have incorporated the tabulated realistic equation of state (EOS) 
based on the relativistic mean field theory \citep{shen98}
instead of the ideal gas EOS assumed in the original code. 

Spherical coordinates are used. No equatorial symmetry is assumed and
the computation domain covers the whole meridian section with 60 angular 
mesh points, except for a model in which we adopted 120 angular grid points. 
Since the latter model did not produce any significant difference from other models, 
we will report in the following the results obtained from the models with 60 angular 
mesh points.  We use 300 radial mesh points to cover
$r_{\rm in} \leq r \leq r_{\rm out} = {2000} ~{\rm km}$,
where $r_{\rm in}$ is the inner boundary and chosen to be the radius of neutrino sphere,
$r_{\nu}$, defined later.

\subsection{Neutrino Absorptions and Emissions by Free Nucleons}

$Q_{\rm E}$ and $Q_{\rm N}$ in Eqs.~(\ref{eq:energy}) and (\ref{eq:ye_flow}) 
are concerning the absorptions and emissions of electron-type neutrino $\nu_{\rm e}$
and anti-neutrino $\bar{\nu}_{\rm e}$ on free nucleons. We summarize here the expressions
for these terms together with the approximations we employ in this paper.

The neutrino absorptivity on free neutron
($\nu_{\text{e}} + \text{n} \rightarrow \text{e}^{-} + \text{p}$)
is written as follows \citep{bruenn_85,rampp_janka_02},
\begin{equation}
 \kappa(\epsilon)
  = \mathcal{G} \eta_{\text{np}} (3g_{\text{A}}^2 + g_{\text{V}}^2)
  \left[ 1 - F_{\text{e}}(\epsilon + \Delta) \right]
  (\epsilon + \Delta)
  \sqrt{(\epsilon + \Delta)^{2} - m_{\text{e}}^2 c^{4}},
\end{equation}
where
$m_{\rm e}$, $c$, $\epsilon$
are electron mass, speed of light, and neutrino energy, respectively.
$\mathcal{G}$ is defined as follows,
\begin{equation}
 \mathcal{G} = \frac{\sigma_{0}}{4m_{\text{e}}^{2}c^{4}},
\end{equation}
where the characteristic cross section of weak interaction, $\sigma_{0}$, is
given as
\begin{equation}
 \sigma_{0}
  = \frac{4(m_{\text{e}}c^{2}G_{\text{F}})^{2}}{\pi(\hbar c)^{4}}
  = 1.761\cdot 10^{-44}~{\rm cm}^{2}.
\end{equation}
In the above equation, $G_{\rm F}$ is the Fermi constant.
$\Delta$ is the difference of rest mass energies of
neutron and proton;
\begin{equation}
 \Delta = m_{\text{n}}c^{2} - m_{\text{p}}c^{2}.
\end{equation}
The nucleon form factors for the vector and axial vector currents are
set to be $g_{\rm V} = 1$ and $g_{\rm A} = 1.23$, respectively. 
The degeneracy factor, $\eta_{\rm np}$, is defined by
\begin{equation}
 \eta_{\text{np}}
 = 2 \int\frac{d^{3}{p}}{(2\pi\hbar c)^{3}}
 F_{\text{n}}(\epsilon)[1 - F_{\text{p}}(\epsilon)].
 \label{eq:eta_np}
\end{equation}
$F_{i}$ is the Fermi-Dirac distribution function for species $i$;
\begin{equation}
 F_{i}(\epsilon_{i})
  = 1/\left[ 1+\exp((\epsilon_{i}-\mu_{i})/k_{\rm B}T) \right],
\end{equation}
where $k_{\rm B}$, $T$, $\mu_{i}$ are Boltzmann constant, temperature,
and chemical potential including rest mass, respectively. The emissivity of $\nu_{\rm e}$
can be obtained from the absorptivity by detailed balance as
\begin{equation}
 j(\epsilon)
  = \exp\left[
	 -\left(
	   \epsilon - (\mu_{\text{p}} - \mu_{\text{n}} + \mu_{\text{e}})
	  \right)/k_{\text{B}}T
	\right]
  \kappa(\epsilon).
  \label{eq:nue-emit}
\end{equation}

The absorptivity of anti-neutrino on free proton
($\bar{\nu}_{\text{e}} + \text{p} \rightarrow \text{e}^{+} + \text{n}$)
is similarly represented by
\begin{equation}
 \kappa(\epsilon)
  = \mathcal{G} \eta_{\text{pn}} (3g_{\text{A}}^2 + g_{\text{V}}^2)
  \left[ 1 - F_{\text{e}^{+}}(\epsilon - \Delta) \right]
  (\epsilon - \Delta)
  \sqrt{(\epsilon - \Delta)^{2} - m_{\text{e}}^2 c^{4}}
  \Theta(\epsilon - \Delta - m_{\text{e}} c^{2}),
\end{equation}
where $\Theta(\epsilon)$ is the step function;
\begin{equation}
 \Theta(\epsilon) =
  \left\{
  \begin{array}{ll}
   1 & \quad\mbox{for $\epsilon \geq 0$}\\
   0 & \quad\mbox{for $\epsilon < 0$}
  \end{array}
  \right. .
\end{equation}
The degeneracy factor, $\eta_{\rm pn}$, is defined by
exchanging the contributions from proton and neutron
in Eq.~(\ref{eq:eta_np}).
The emissivity is obtained again by the detailed balance just like 
for Eq.~(\ref{eq:nue-emit}) together with the relation of 
$\mu_{{\rm e}^{+}} = -\mu_{\rm e}$;
\begin{equation}
 j(\epsilon)
  = \exp\left[
	 -\left(
	   \epsilon - (\mu_{\rm n} - \mu_{\rm p} - \mu_{\rm e})
	  \right)/k_{\rm B}T\right] \kappa(\epsilon).
\end{equation}

Using the absorptivities and emissivities given above,
$Q_{\text{E}}$ and $Q_{\text{N}}$ can be expressed as follows,
\begin{equation}
 Q_{\text{E}}
  =-\frac{4\pi c}{(2\pi\hbar c)^{3}}
  \int_{0}^{\infty}\epsilon^{3} d\epsilon
  \left[
   j(\epsilon) - \left(j(\epsilon) + \kappa(\epsilon)\right) f(r, \epsilon)
  \right],
\end{equation}
\begin{eqnarray}
 Q_{\text{N}}
  = i \frac{m_{\text{B}}}{\rho} \frac{4\pi c}{(2\pi\hbar c)^{3}}
  \int_{0}^{\infty}\epsilon^{2} d\epsilon
  \left[
   j(\epsilon) - \left(j(\epsilon) + \kappa(\epsilon)\right) f(r, \epsilon)
  \right], \\
 \left\{
  \begin{array}{ll}
   i=-1  &\quad\mbox{(for electron-type neutrino)} \\
   i= 1  &\quad\mbox{(for electron-type antineutrino)}
  \end{array}
 \right. ,\nonumber
\end{eqnarray}
where $f(r, \epsilon)$ is the distribution function of neutrino.
These rates are calculated
for $\nu_{\rm e}$ and $\bar{\nu}_{\rm e}$ separately 
and summed for the source terms
in Eqs.~(\ref{eq:energy}) and (\ref{eq:ye_flow}).

Since we deal with the optically thin region outside the neutrino sphere in this paper,
we do not have to solve the transport equation for neutrinos. We assume that 
the neutrino distribution functions are approximated by the Fermi-Dirac distribution 
with a vanishing chemical potential; 
\begin{equation}
 f(r, \epsilon)
  = \frac{1}{1+\exp(\epsilon/k_{\text{B}}T_{\nu})} \cdot
  \frac{1 - \sqrt{1 - ( r_{\nu}/r )^{2}}}{2\pi},
\end{equation}
where the geometrical factor is taken into account for normalization. 
Note that although the angular dependence in the above distribution function is 
assumed to be isotropic, it is entirely irrelevant for the absorption and emission rates. 
We further assume for simplicity in the following that the luminosity $L_{\nu}$, 
temperature $T_{\nu}$ and neutrino sphere $r_{\nu}$ are related by the following equation,
\begin{equation}
 L_{\nu}
  = \frac{7}{16}\sigma T_{\nu}^{4} \cdot 4\pi r_{\nu}^{2},
\end{equation}
where $\sigma$ is the Stefan-Boltzmann constant.

\subsection{Initial and Boundary Conditions}

In order to clearly see the linear growths of the instability,
it is critically important to use a well-defined unperturbed state 
as an initial condition for simulations. For this purpose,
we employ the spherically symmetric steady accretion flows
through the standing shock wave in this paper. Following \citet{yamasaki}, we solve 
the time-independent hydrodynamical equations from the shock front down to the inner boundary,
\begin{equation}
 4\pi r^{2}\rho v_{r} = \dot{M}  \label{eq:steady1},
\end{equation}
\begin{equation}
 v_{r}\frac{dv_{r}}{dr}
  +\frac{1}{\rho}\frac{dP}{dr} + \frac{GM_{\rm in}}{r^{2}} = 0,
\end{equation}
\begin{equation}
 v_{r}\frac{de}{dr}
  - \frac{P}{\rho^{2}}v_{r}\frac{d\rho}{dr}
  = \frac{Q_{\text{E}}}{\rho},
\end{equation}
\begin{equation}
 v_{r}\frac{dY_{\rm e}}{dr} = Q_{\rm N},
  \label{eq:steady4}
\end{equation}
where $\dot{M}$ and $v_{r}$
are the mass accretion rate and radial velocity, respectively.
Given the post-shock values for density, radial velocity, entropy and 
electron fraction ($\rho_{\text{post}}$, $v_{\text{post}}$, $s_{\text{post}}$, 
${Y_{\text{e}}}_{\text{post}}$),
the shock radius is determined so that the density obtained at the inner boundary 
should agree with the fixed value, $\rho_{\rm in} = 10^{11}~\mbox{g/cm}^{3}$.

The post-shock values are calculated from the corresponding 
pre-shock values by the Rankine-Hugoniot relations. The
pre-shock values, on the other hand, are obtained from the mass accretion rate, the 
outer boundary conditions for entropy and electron fraction 
($s_{\text{out}}$, ${Y_{\text{e}}}_{\text{out}}$) and the assumption that
matter falls freely with no interaction with neutrinos outside the shock;
\begin{equation}
 v_{\text{pre}} = -\sqrt{2GM_{\rm in}/R_{\text{s}}},
\end{equation}
\begin{equation}
 \rho_{\text{pre}} = -\frac{\dot{M}}{4\pi r^{2}v_{\text{pre}}},
\end{equation}
\begin{equation}
 s_{\text{pre}} = s_{\text{out}},
\end{equation}
\begin{equation}
 {Y_{\text{e}}}_{\text{pre}} = {Y_{\text{e}}}_{\text{out}},
\end{equation}
where $R_{\rm s}$ is the shock radius. 
The entropy and the electron fraction at the outer boundary
are assumed to be $s_{\rm out} = 3 ~k_{\rm B}\mbox{/baryon}$ 
and ${Y_{\rm e}}_{\rm out} = 0.5$, respectively.

In the numerical simulations, axisymmetry but no equatorial symmetry is assumed. 
At the outer boundary, we adopt the fixed boundary condition consistent with the 
unperturbed state. On the other hand, the free-flow-in boundary condition is used at the 
inner boundary. A realistic tabulated EOS by \citet{shen98} is used both for the simulations
and for the preparation of the initial conditions. 

For all the models investigated in this paper, the mass accretion rate and the mass of central object are fixed
to be $\dot{M} = 1~M_{\odot}\mbox{/s}$ and $M_{\rm in} = 1.4~M_{\odot}$, respectively. 
The neutrino temperatures are also fixed to $T_{\nu_{\text{e}}} = 4$ MeV and 
$T_{\bar{\nu}_{\text{e}}} = 5$ MeV, the typical values in the post-bounce phase. 
On the other hand, we systematically vary the neutrino luminosity within the range 
of $L_{\nu} = \mbox{3.0--7.0}\cdot 10^{52}~\mbox{erg/s}$ for different models. 
Note that these values are constant in time for each model. This is necessary
to realize the steady unperturbed states.  

To study the instability, we add angular-dependent perturbations, $\delta v_{r}(r, \theta)$, 
to the radial velocity initially;
\begin{equation}
 v_{r}(r, \theta)
  = v_{r}^{\text{1D}}(r)
  + \delta v_{r}(r, \theta),
\end{equation}
where $v_{r}^{1D}$ is the spherically symmetric unperturbed velocity. 
The $\ell=1$ ($\delta v_{r}(r, \theta) \propto \cos \theta$) single-mode perturbation and 
the random multi-mode perturbation are investigated.  
In the former case, the perturbation amplitude is set to be 1\% of the unperturbed velocity.
For the latter case, the amplitude is less than 1\% for each radial direction. 
In both cases, the assumed initial amplitudes are found to be small enough to observe clearly the 
linear growths of the instability.

\section{Results\label{sec3}}

\subsection{Basic Features}

\begin{figure}
\epsscale{1.0}
\plotone{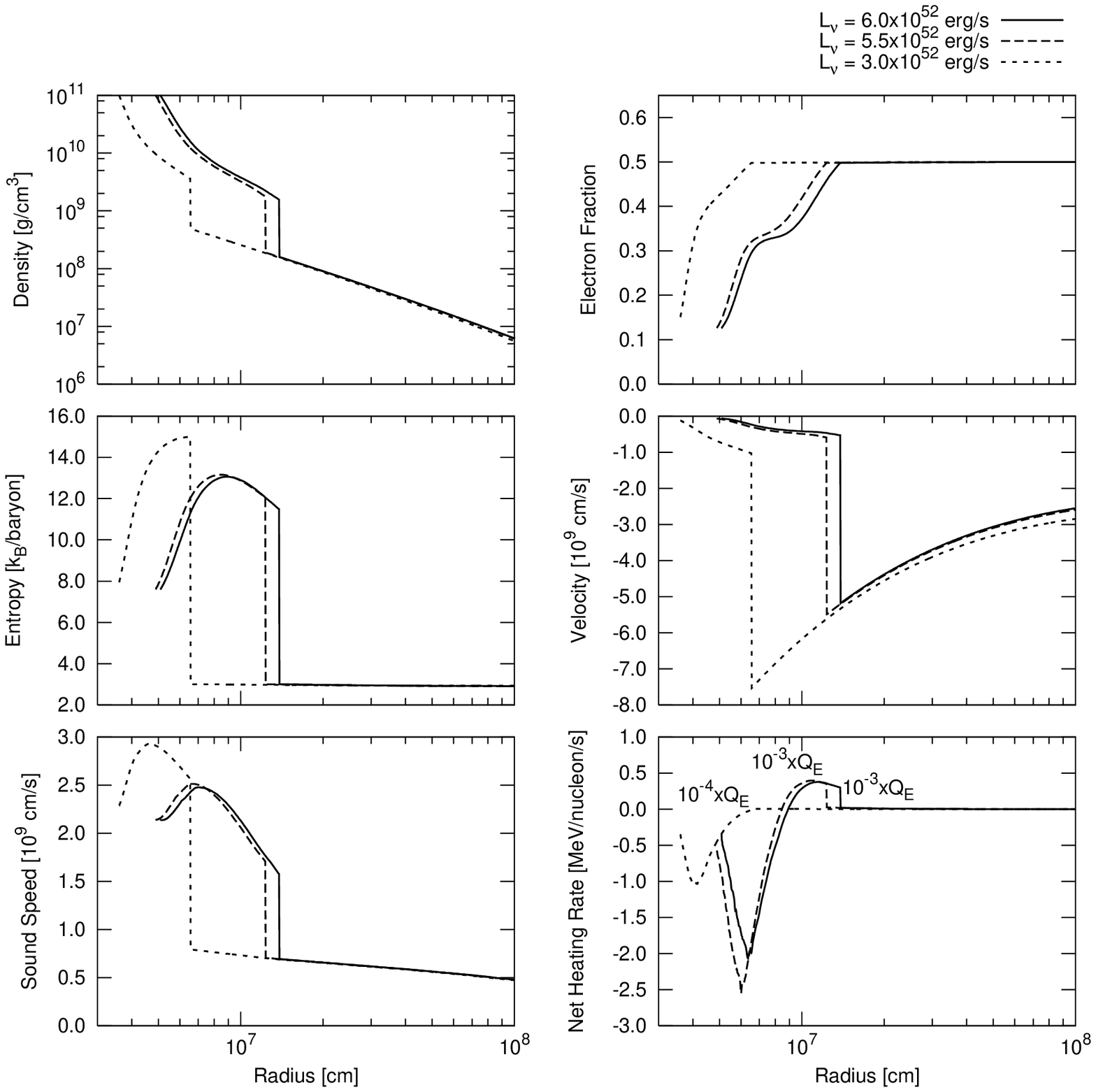}
\caption{Physical quantities in unperturebed spherically symmetric accretion flows.
Solid, dashed, and dotted lines represent the solutions
for $L_{\nu} = 6.0\cdot 10^{52}$, $5.5\cdot 10^{52}$,
and $3.0\cdot 10^{52}$ erg/s,
respectively.
}
\label{fig:initial_profiles}
\end{figure}

The unperturbed spherically symmetric steady accretion flows are displayed
in Fig.~\ref{fig:initial_profiles} for $L_{\nu} = 3.0, 5.5, 6.0\cdot 10^{52}~\mbox{erg/s}$.
As $L_{\nu}$ increases, the width between the shock and 
the neutrino sphere, $w_{\rm s} = R_{\rm s} - r_{\nu}$,
becomes larger with both the neutrino sphere 
and the shock radius getting larger.
In the cases of $L_{\nu} = 5.5, 6.0\cdot 10^{52}~\mbox{erg/s}$,
the region with a negative entropy-gradient is formed behind the shock,
where the net heating rate is positive.
The supernova core is convectively unstable in this region.
In the case of $L_{\nu} = 3.0\cdot 10^{52}~\mbox{erg/s}$, on the other hand, 
the net heating rate is always negative over the whole region from the shock 
down to the neutrino sphere, since the shock radius is smaller and the temperature
is higher, implying more efficient coolings. As a result, there is no region with 
a negative entropy-gradient. Hence we can see in this case the instability 
solely caused by SASI. It is also noted that this model should be similar to 
the models considered by \citet{blondin_05} with only simple cooling terms 
taken into account.

We use the profiles in Fig.~\ref{fig:initial_profiles} as initial conditions for
hydrodynamical simulations in 2D. Since they are obtained by solving 
Eqs.~(\ref{eq:steady1})--(\ref{eq:steady4}), a slight inconsistency is 
inevitably introduced by the mapping. For example, the shock is smoothed over 
a few computational grid points with the artificial viscosity in the simulations while 
the initial conditions satisfy the Rankine-Hugoniot relation across the infinitesimal distance.
As a result, transient oscillations are commonly found before they are damped away. 
In fact, we have done spherically symmetric computations for the unperturbed initial 
conditions using the same dynamical code before doing 2D simulations with perturbations. 
The purpose is two-fold. First, we can obtain this way "the steady states" for the dynamical code, 
which, as we have emphasized repeatedly, is critically important to see the linear growth of 
the instability. These simulations are also meant for the confirmation of
the stability of the initial conditions against radial perturbations predicted by \citet{yamasaki}.

\begin{figure}
\epsscale{1.0}
\plotone{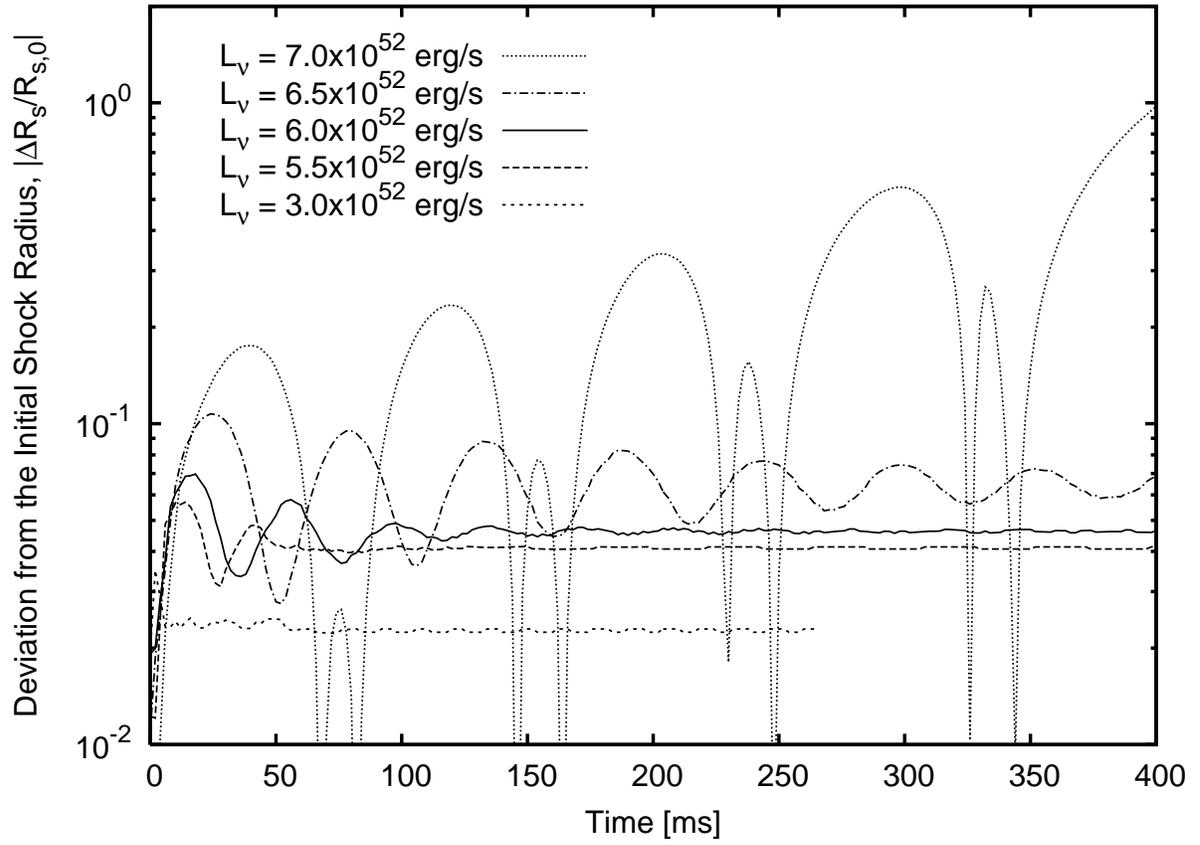}
\caption{Temporal evolutions of shock radius in 
the spherically symmetric simulations for the models given 
in Fig.~\ref{fig:initial_profiles} together with for $L_{\nu} = 6.5, 7.0\cdot 10^{52}~\mbox{erg/s}$. The relative deviation from the initial value is plotted}
\label{fig:shock_radius_1D}
\end{figure}

The temporal evolutions of the shock radii in the spherically symmetric simulations
are shown in Fig.~\ref{fig:shock_radius_1D} for the three cases in Fig.~\ref{fig:initial_profiles}
together with for $L_{\nu} = 6.5, 7.0\cdot 10^{52}~\mbox{erg/s}$.
The oscillations of shock radii are particularly clear in the early phase.
The oscillation frequency is found to be inversely proportional to the neutrino luminosity.
For the neutrino luminosity less than $7.0\cdot 10^{52}~\mbox{erg/s}$,
the oscillations are gradually damped and the shock 
approaches to the equilibrium radius, which is a bit different from the initial value. 
In the case of $L_{\nu} =7.0\cdot 10^{52}~\mbox{erg/s}$, however, 
the oscillation is slowly amplified and appears to be never settled to equilibrium.
It is noted that this model is expected to be stable according to \citet{yamasaki}.
The apparent discrepancy may be ascribed to the numerical errors inherent to the 
dynamical simulations as mentioned above. It should be also mentioned, however, that the large-amplitude
oscillations as observed in the figure may cause instability even for linearly stable configurations.
Since the main purpose of this paper is to investigate the stability for non-radial perturbations,
we focus in the following only on the models with $L_{\nu} \leq 6.0\cdot 10^{52}~\mbox{erg/s}$, 
in which the radial oscillations are damped within $\sim 100~\mbox{ms}$. 
After these transient feateures are sufficiently settled, we add non-radial velocity perturbations 
explained in the previous section to the profiles obtained this way and start 2D simulations.

\begin{figure}
\epsscale{1.0}
\plottwo{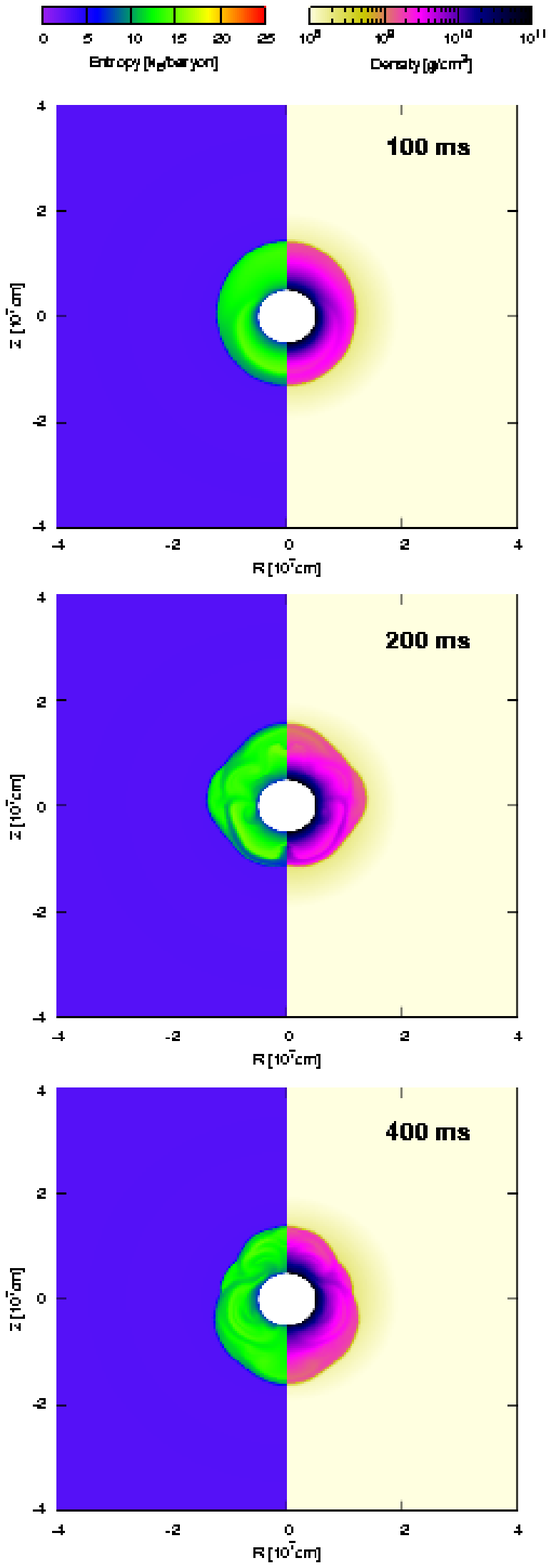}{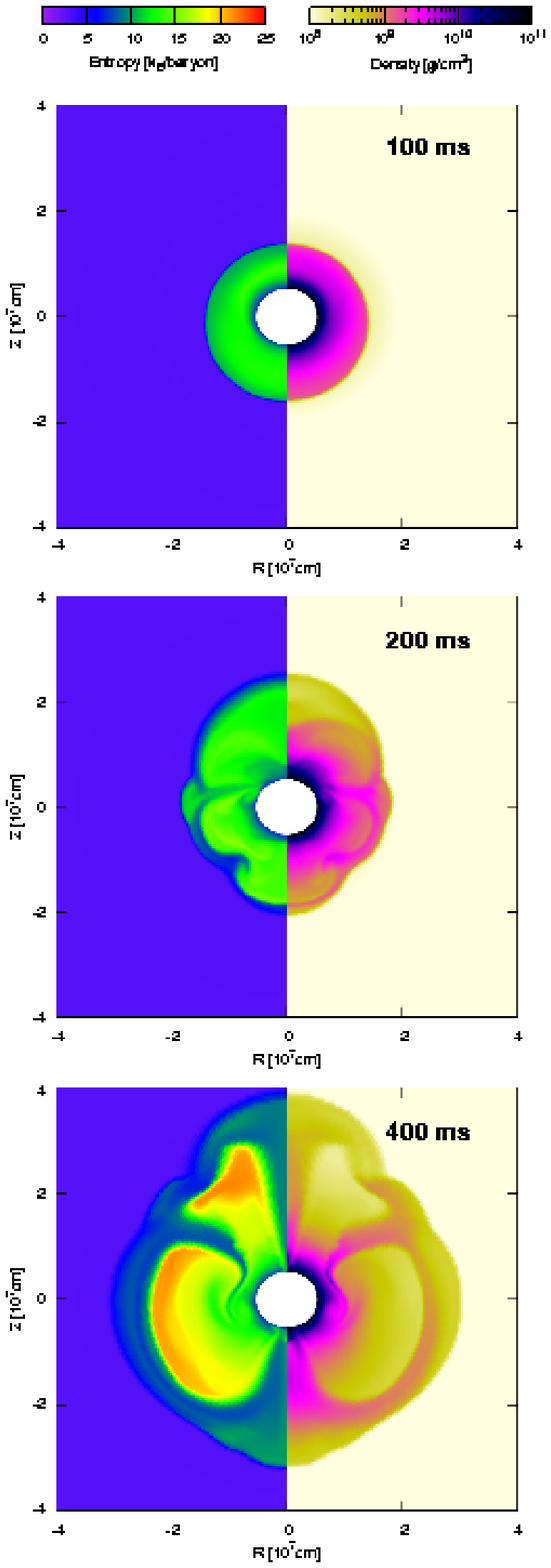}
\caption{Entropy- (the left half of each panel) and density- (the right half) 
distributions in the meridian section for 1\% of the $\ell = 1$ single-mode 
velocity perturbation. $L_{\nu} = 5.5\cdot 10^{52}$ erg/s is assumed for the
left panels and $L_{\nu} = 6.0\cdot 10^{52}$ erg/s is for the right panels.
}
\label{fig:entrdens_Ln55_Ln60}
\end{figure}

Figure~\ref{fig:entrdens_Ln55_Ln60} shows in the meridian section the distributions 
of entropy (the left half of the panel) and density (the right half) for the models 
with $L_{\nu} = 5.5, 6.0\cdot 10^{52}~\mbox{erg/s}$ after 1\%  of the $\ell = 1$ single-mode velocity 
perturbation is added. For both models, we observe the growth of the perturbations. 
In the case of $L_{\nu} = 5.5\cdot 10^{52}~\mbox{erg/s}$,
the shock surface is deformed at first by the increasing amplitude of the non-radial mode and then 
begins to oscillate with a large amplitude. In the case of $L_{\nu} = 6.0\cdot 10^{52}~\mbox{erg/s}$ (right panels),
on the other hand, in addition to the oscillations of the shock surface, we observe the substantial increase
of the average shock radius as the time passes. In fact, after $t=400~\mbox{ms}$, the shock radius 
continues to increase and appears to produce an explosion.
Since the model is stable against radial perturbations as mentioned above, the non-radial instability 
and the neutrino heating therein are responsible for the explosion. We think that this is a reconfirmation 
of the claim that the instability, whatever the cause, behind the shock is helpful for the shock revival.

In order to analyze further in detail the evolution of the instability, we conduct a mode analysis as follows. 
The deformation of the shock surface is decomposed into the spherical harmonic components;
\begin{equation}
 R_{\rm s}(\theta) = \sum_{\ell=0}^{\infty}a_{\ell}
  \sqrt{\frac{2\ell+1}{4\pi}}P_{\ell}(\cos\theta).
\end{equation}
Since the system is axisymmetric, only $m=0$ harmonics, nothing but Legendre polynomials, show up. 
The coefficients, $a_{\ell}$, can be calculated by the orthogonality of the Legendre polynomials;
\begin{equation}
 a_{\ell} = \frac{2\ell+1}{2}\int_{-1}^{1}R_{\rm s}(\theta)
  P_{\ell}(\cos\theta)d\cos\theta.
\end{equation}
The position of the shock surface, $R_{\rm s}(\theta)$, is estimated from
the iso-entropic surface of $s=5$ in the following analyses.

\begin{figure}
\epsscale{1.0}
\plotone{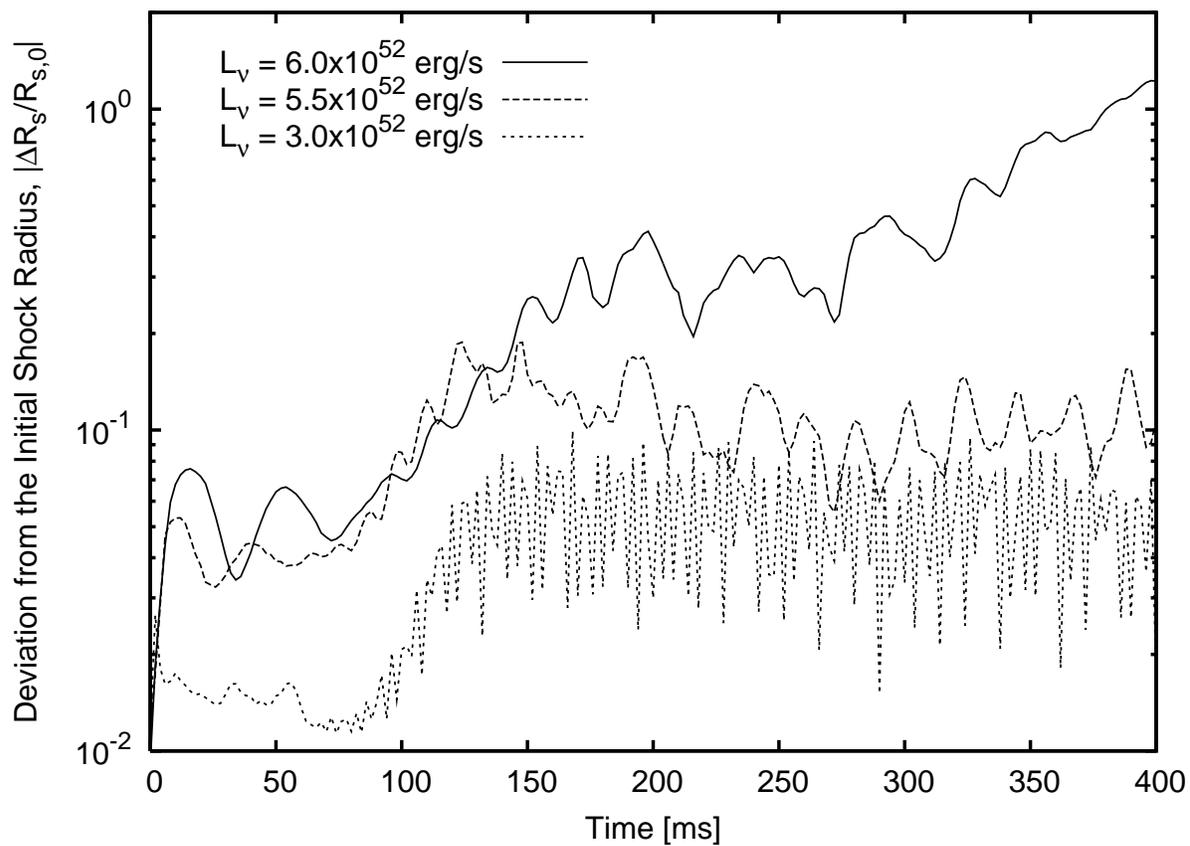}
\caption{Temporal evolutions of the $\ell = 0$ component in the spherical harmonic decompositions. 
The relative deviation from the initial value is plotted.}
\label{fig:shock_radius}
\end{figure}

The temporal evolutions of the average shock radius are obtained from the $\ell = 0$ component in 
the above decomposition and $|\Delta R_{\rm s}/R_{\rm s,0}|=|(a_{0}-R_{\rm s,0})/R_{\rm s,0}|$
are plotted in Fig.~\ref{fig:shock_radius}, where $R_{\rm s,0}$ is the initial shock radius.  
As can be seen from the figure, after the shock surface oscillates for about 100~ms 
as in the symmetric simulations, it begins to increase. 
For the models with $L_{\nu} = 3.0, 5.5\cdot 10^{52}~\mbox{erg/s}$,
the shock surface is settled in $\sim50$~ms to a quasi-steady state with a radius $\sim10$\% larger 
than the initial value, and keeps oscillating around it thereafter. 
In the case of $L_{\nu} = 6.0\cdot 10^{52}~\mbox{erg/s}$, on the other hand, 
a continuous increase of the average shock radius is found, which will eventually lead to the explosion
mentioned above. It is emphasized that even for the model
with no negative entropy-gradient ($L_{\nu} = 3.0\cdot 10^{52}~\mbox{erg/s}$),
the exponential increase of the shock radius occurs during the same period.
This demonstrates clearly that the instability has nothing to with the convection 
at least for this model. The relative contributions of SASI and convection for other models will be 
discussed in the next section.

\subsection{Linear and Nonlinear Growths of Instability}

In this section, we discuss both the linear and nonlinear growths of the instability in greater detail.
As reported in \citet{foglizzo01,foglizzo02,blondin_03}, the most remarkable feature of SASI is the dominance of 
the modes with $\ell=1, 2$ in the instability. Hence we start with the analysis of the models 
with $\ell=1$ single-mode velocity perturbation imposed initially. 

\begin{figure}
\epsscale{1.0}
\plotone{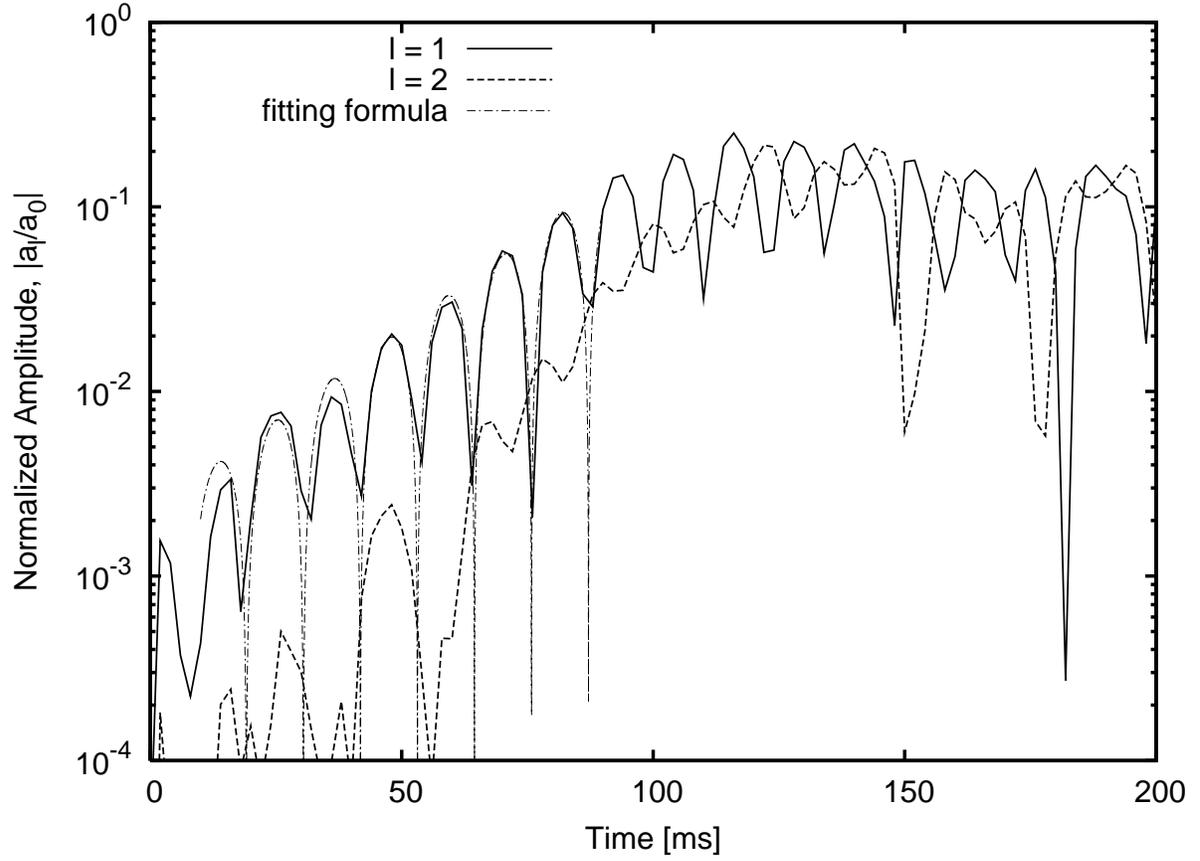}
\caption{Temporal evolutions of the normalized amplitudes
of the $\ell=1, 2$ modes for the model with $L_{\nu} = 5.5\cdot 10^{52}$ erg/s.
The dot-dashed line represents the fitting in the linear phase.}
\label{fig:fitting_Ln55_steady}
\end{figure}

Fig.~\ref{fig:fitting_Ln55_steady} shows the temporal evolutions of the
$\ell=1$ and $\ell=2$ modes. Although the result for $L_{\nu} = 5.5\cdot 10^{52}~\mbox{erg/s}$
is plotted as a reference case, the qualitative feature is common to other models. 
As can be clearly seen, the evolution is divided into two phases. The initial phase lasting for 
$\sim 100$~ms represents the linear phase, where the amplitude of each mode grows exponentially. It is noted
that even for the single-mode perturbation, the second harmonics is generated by the nonlinear coupling 
and grows exponentially also. In the linear phase, however, the second harmonics is always much smaller than 
the fundamental $\ell =1$ mode. After $\sim 100$~ms, the amplitude of the fundamental mode is saturated 
at $\sim 10$\% level and the nonlinear phase starts. It is found that soon after the nonlinear phase begins,  
the amplitude of the second harmonics becomes comparable to the fundamental mode. This corresponds to 
the instability found in the simulations by \citet{blondin_03}. 

In order to obtain the linear growth rates for the fundamental mode ($\ell=1$),
we fit the amplitude of the mode, $a_{1}(t)$, with the following expression,
\begin{equation}
 a_{1}(t) = A\exp(\gamma t)\sin(\omega t + \delta),
\end{equation}
where $\gamma$ and $\omega$ are
the exponential growth rate and the characteristic oscillation frequency
in the linear phase, respectively. The least square fitting is done for these parameters
and the overall normalization, $A$, and the phase, $\delta$.
The dash dotted line in Fig.~\ref{fig:fitting_Ln55_steady}
represents the result. The obtained values of $\gamma$ and $\omega$ are listed
in Table~\ref{table:growth_rate} for the models with different $L_{\nu}$.

According to \citet{foglizzo01,foglizzo02}, the instability is produced by
the cycle of the inward advection of the velocity and entropy
fluctuations and the outward propagation 
of the pressure fluctuations (see also \citet{blondin_03}). If this is correct, the characteristic 
oscillation frequency reflects the the cycle period, 
\begin{equation}
\omega = 2 \pi \cdot \left\{\int_{r_{\nu}}^{R_{\rm s}} \!\!\!\!dr \left[ \frac{1}{c_{\rm s}} 
+ \frac{1}{v_r}\right]\right\}^{-1},
\end{equation}
where $c_{\rm s}$ is the sound velocity. The values of $\omega$ estimated this way are also given 
in Table~\ref{table:growth_rate}. They are found to agree quite well with the numerical results. 
This is in contrast with the recent claim by \citet{blondin_05} that the propagation of
pressure perturbations is responsible for the instability. Their simplified treatment of the
cooling terms may be responsible for the difference. Our simulations indicate clearly that 
the time scale associated with the advection plays an important role in the instability. 
We can obviously find the $L_{\nu}$-dependence of the oscillation frequency, $\omega$, 
which can be understood as follows. As also presented in Table~\ref{table:growth_rate}, 
the width between the shock and the neutrino sphere,
$w_{\rm s} = R_{\rm s,equil} - r_{\nu}$, 
becomes larger as the luminosity increases. As a result, the oscillation frequency is expected to 
become lower since the fluctuations traverse longer distances in the cycle.

The growth rate, $\gamma$, on the other hand, has little dependence on the luminosity as shown 
in Table~\ref{table:growth_rate}, As mentioned already, the instability occurs even in the case of 
$L_{\nu} = 3.0\cdot 10^{52}~\mbox{erg/s}$, where the negative entropy-gradient is not formed and 
the structure is stable for convection. The interesting thing here is the fact that the linear growth rates are
not so different between the models with and without a negative entropy-gradient.
This suggests that SASI plays a dominant role in driving the non-radial motions even when 
the convection is also expected to occur. This issue will be further discussed later in this section again.

The simulations discussed so far have been done with the initial velocity perturbation including
only the fundamental mode ($\ell=1$). However, other modes with $\ell \geq 2$ also develop rapidly 
from $\sim 50~\mbox{ms}$. After $\sim 100~\mbox{ms}$, the amplitude of $\ell=2$ mode becomes 
the same order as that of the fundamental mode, which has already been saturated by this time. This marks the
beginning of the nonlinear phase. In fact, the $\ell=2$ mode is also soon saturated. 
This transition from linear to nonlinear phase corresponds to the time of the rapid increase of 
the average shock radius shown in Fig.~\ref{fig:shock_radius}. Since the average radius is nothing but 
the $\ell=0$ mode, this can be interpreted as a result of the nonlinear coupling of this mode with the 
fundamental mode and the ensuing saturation. As will be mentioned later, since the expansion of the shock 
is crucial to trigger the shock revival and eventual explosion, it is important that the neutrino luminosity 
is sustained for $\sim 100~\mbox{ms}$, the typical saturation time for SASI.

\begin{figure}
\epsscale{1.0}
\plottwo{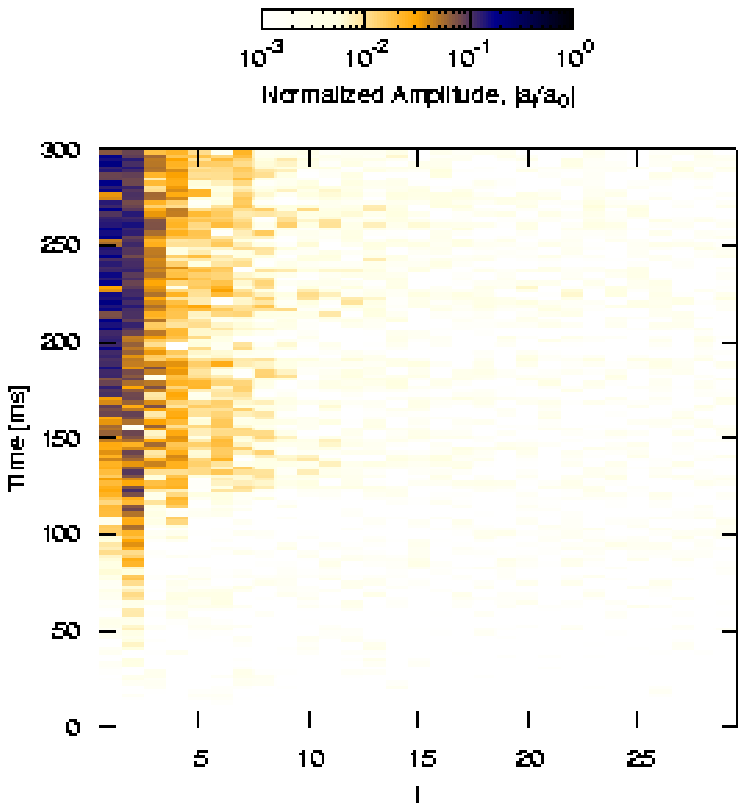}{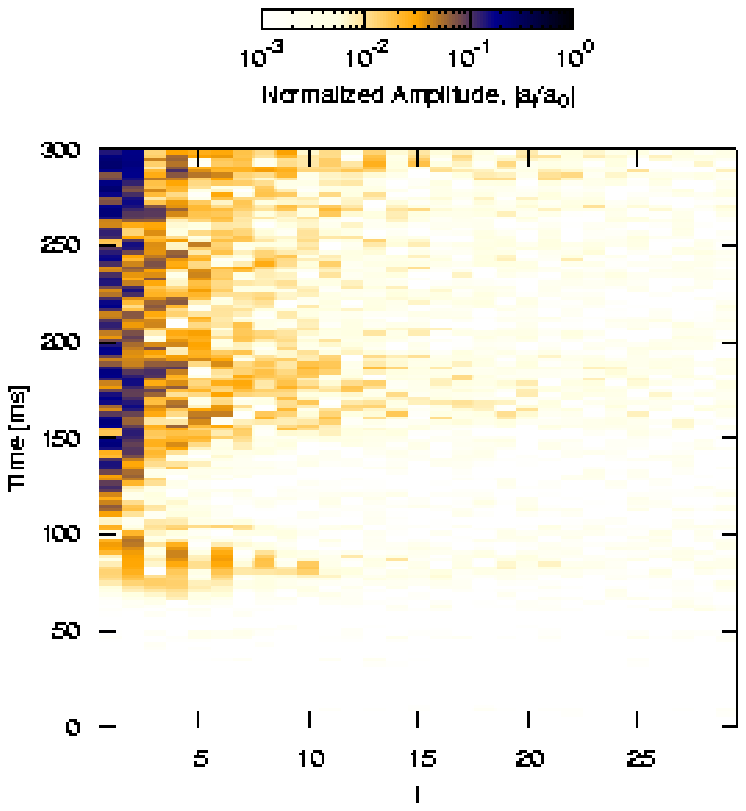}
\caption{Temporal evolutions of the spectra in the spherical harmonics decomposition for the models
with $L_{\nu} = 3.0\cdot 10^{52}$ erg/s (left panel) and with $L_{\nu} = 5.5\cdot 10^{52}$ erg/s (right panel).
The random multi-mode velocity perturbations are initially added.
}
\label{fig:spectra_rand_Ln30_Ln55}
\end{figure}

Next we discuss the models with the random multi-mode velocity perturbations.
So far we confirmed that the $\ell=1$ mode is indeed unstable to SASI. The models are
meant to see which mode is dominant. In so doing, we also study the influence of the 
existence of a negative entropy-gradient. In Fig.~\ref{fig:spectra_rand_Ln30_Ln55},
the temporal evolution of the spectrum of the spherical harmonics. The spherically symmetric
component, the $\ell=0$ mode, is omitted in the figure. The cases without and with 
a negative entropy-gradient are shown in the left and right panels, respectively. 
It is obvious that the modes with small $\ell$'s, especially those with $\ell=1, 2$,
grow rapidly in the linear regime ($t \lesssim 100~\mbox{ms}$).
This is particularly the case for the model without a negative entropy-gradient
($L_{\nu} = 3.0\cdot 10^{52}~\mbox{erg/s}$)
and the growths of the modes with $\ell>10$ are negligibly small.
With a negative entropy-gradient ($L_{\nu} = 5.5\cdot 10^{52}~\mbox{erg/s}$),
the broadening of spectra to larger $\ell$ modes is observed although the dominance of 
smaller $\ell$ modes can be still found. The convective instability may enhance
the growth of higher harmonics in the linear phase. The similarity of the two cases suggest 
again that SASI is dominant over the convection even when the latter is operating.

Recently \citet{foglizzo05} discussed the linear stability for convection in the 
accretion flows in the supernova core. They found that the classical criterion for convection, 
that is, the negative entropy-gradient is not sufficient for the accretion flows, since the limited 
time is available for growth. Although the classical convection has greater linear-growth rates
for modes with larger wave numbers, they claimed that there are minimum
($k_{\rm min}$) and 
maximum ($k_{\rm max}$) wave numbers for unstable modes in the accretion flows, and that 
the growth rates are also modified. The important parameter is the ratio of the advection time 
through the gain region divided by the local timescale of buoyancy, $\chi$, given in Eq.~(40) in their paper.
Applying their formula to our models, we obtain 
$\chi = 4 \, ... \,  7$ for $L_{\nu} = 5.5\,  ... \, 6.5\cdot 10^{52}~\mbox{erg/s}$, 
with larger values for greater luminosities. Hence, the initial configurations are unstable against convection for these
models, since the criterion, $\chi > 3$ is satisfied. The minimum and maximum
wave numbers estimated in our models are $k_{\rm min}=2 \, ... \, 6\cdot 10^{-8}$~cm$^{-1}$ and
$k_{\rm max}=1\cdot10^{-6}$~cm$^{-1}$, respectively. 
The smaller $k_{\rm min}$ corresponds to larger luminosities. They roughly correspond to the minimum and 
maximum indices in the spherical harmonics, $\ell_{\rm min}= 2 \, ... \,
4$ and $\ell_{\rm max}= 70 \, ... \, 85$, respectively. 
The lower $\ell_{\rm min}$ and larger $\ell_{\rm max}$ are obtained for higher luminosities. These number appear to be 
consistent with the spectrum shown in Fig.~\ref{fig:spectra_rand_Ln30_Ln55} as also inferred from Fig.~5 in \citet{foglizzo05}. Since the classical growth rate of convection is comparable to that of SASI in our models and
the true growth rate of convection will be much smaller than the classical estimation, 
we think that SASI is a dominant driving force for the non-radial motions we observed so far. 

It is also interesting to note that the modes with $\ell=1, 2$ are dominant in the nonlinear regime, 
which begins after $\sim 100$~ms. As clearly seen in the broadening of the spectra in 
Fig.~\ref{fig:spectra_rand_Ln30_Ln55}, various modes are amplified by nonlinear couplings with
the dominant modes in this phase. The spectra are again broader for the model with a negative 
entropy-gradient. In both models, however, the dominance of the modes with $\ell=1, 2$ is remarkable.
This should correspond to the large deformations of shock wave found in the numerical simulations 
by \citet{blondin_03} and \citet{scheck_04}. In order to make clear the reason for
the dominance of these modes in the nonlinear regime, it will be required to study the nonlinear couplings 
of various modes in more detail.

\subsection{Neutrino Heatings in SASI}

\begin{figure}
\epsscale{1.0}
\plottwo{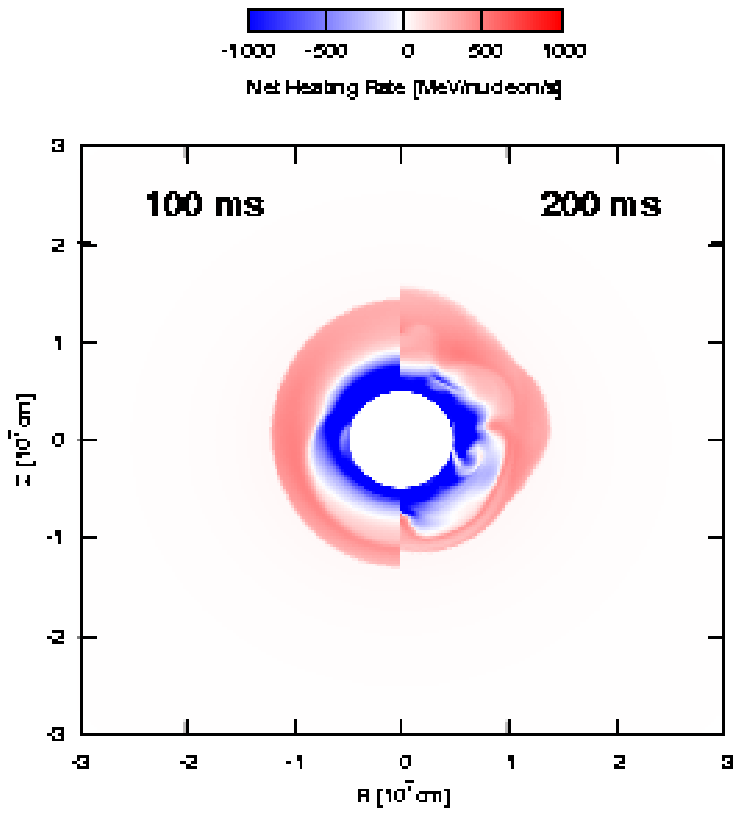}{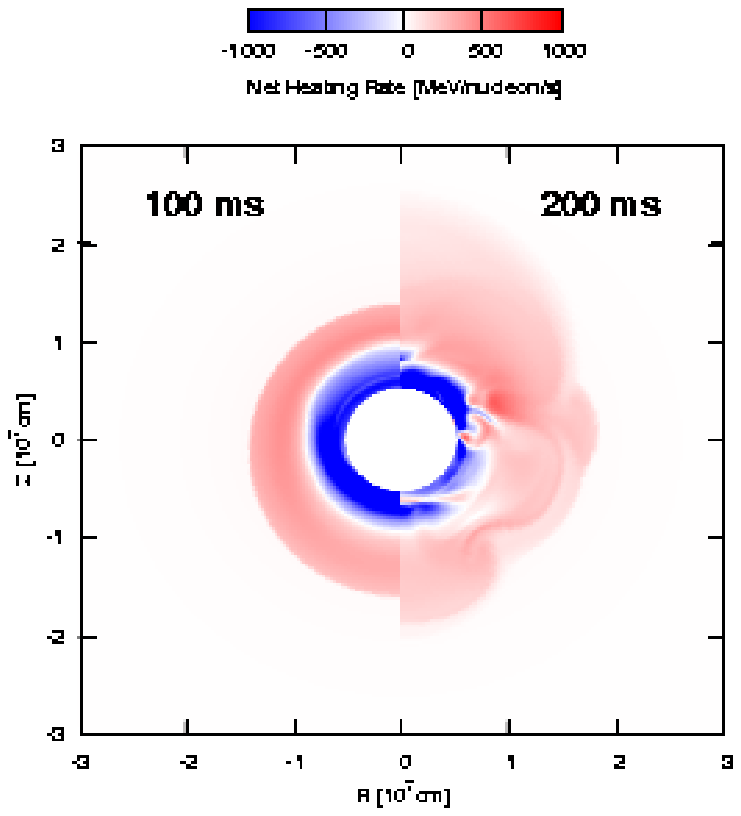}
\caption{Net heating rates in the meridian section for the models
with $L_{\nu} = 5.5\cdot 10^{52}$ erg/s (left panel) and 
$L_{\nu} = 6.0\cdot 10^{52}$ erg/s (right panel). 
1\% of the $\ell=1$ single-mode velocity perturbation is initially added.
The left (right) half of each panel represents the result for 100 (200)~ms.
}
\label{fig:netheat_Ln55_Ln60}
\end{figure}

Finally, we discuss the influence of the instability on the heating of matter by neutrinos.
Figure~\ref{fig:netheat_Ln55_Ln60} shows the distributions of the net heating rates in the
meridian section for the models with $L_{\nu} = 5.5\cdot 10^{52}~\mbox{erg/s}$ (left panel)
and $L_{\nu} = 6.0\cdot 10^{52}~\mbox{erg/s}$ (right panel). In both cases, the shock wave is 
asymmetric with respect to the equator, the characteristics for SASI. 
For $L_{\nu} = 5.5\cdot 10^{52}~\mbox{erg/s}$, however, the average shock radius is 
only slightly larger than that in the unperturbed state. As a result, both the heating and cooling 
region are not much changed during the period from 100~ms to 200~ms. They are just oscillating.
In the case of $L_{\nu} = 6.0\cdot 10^{52}~\mbox{erg/s}$, on the other hand, not only the 
shock radius but also the heating region is getting larger as the time passes. 
It is further seen that the cooling region at 200~ms is smaller than that at 100~ms.
This is caused by the non-radial flows which carry down colder matter more efficiently than 
the spherical flows. The temperature near the inner boundary is lowered as a result. Since the 
cooling rates are roughly proportional to $T^{6}$, the net cooling region is shrunk. This mechanism
has been discussed in the context of convections over the years. It is applicable to SASI as it is. 
This broadening of the heating region and the shrink of the cooling region thrusts the
shock wave outwards further, the positive feed back which eventually leads to the explosion as 
mentioned already.

\begin{figure}
\epsscale{1.0}
\plotone{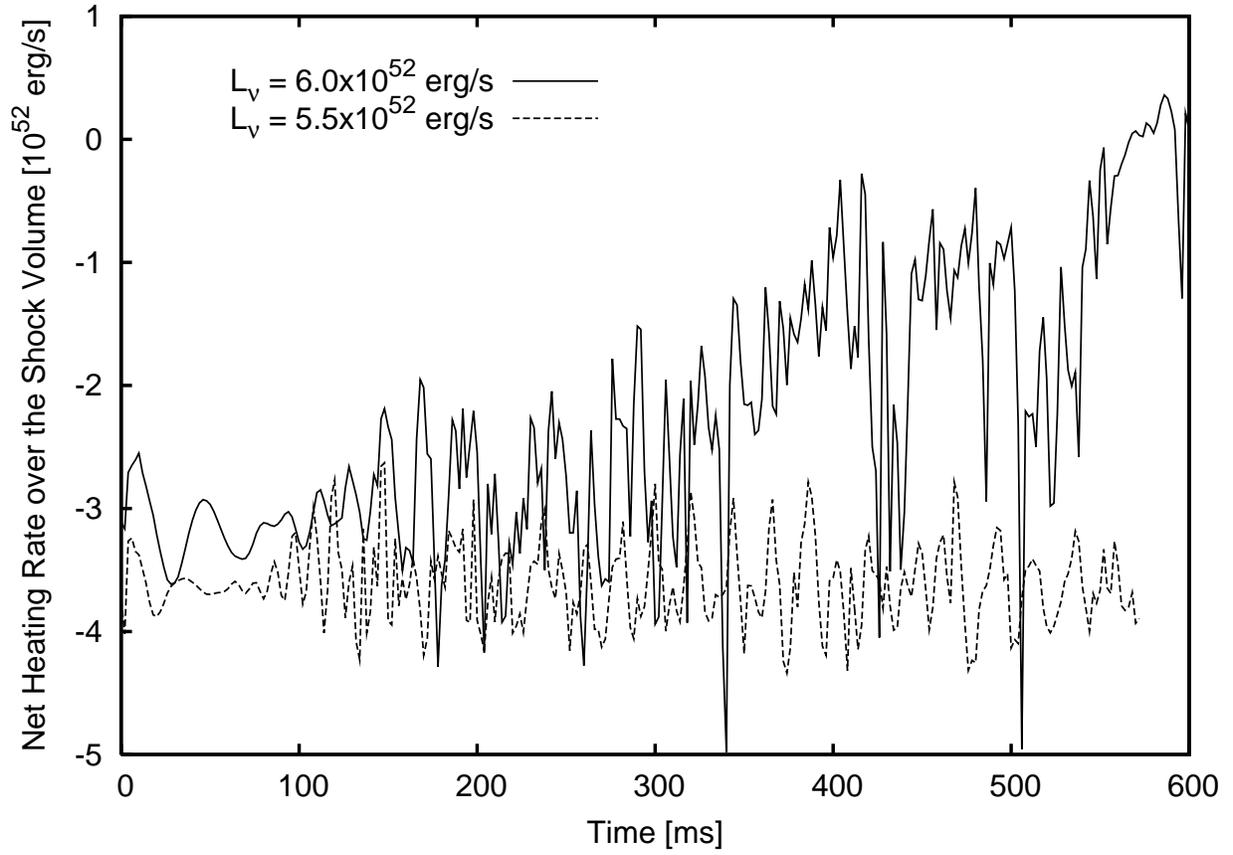}
\caption{Temporal evolutions of the net heating rate integrated over
the region inside the shock wave.
The solid and dashed lines represent the models with
$L_{\nu} = 6.0\cdot 10^{52}~\mbox{erg/s}$ and
$L_{\nu} = 5.5\cdot 10^{52}~\mbox{erg/s}$,
respectively. 1\% of the $\ell=1$ single-mode velocity perturbation is initially added.
}
\label{fig:netheating_Ln60_Ln55}
\end{figure}

This positive feed back cycle is confirmed by the comparison of the temporal evolution of the 
shock radius displayed in Fig.~\ref{fig:shock_radius} and that of the net heating rate integrated 
over the volume inside the shock wave shown in Fig.~\ref{fig:netheating_Ln60_Ln55}.
We can see the gradual and continuous increase of the net heating after $\sim 200$~ms 
coincides with the expansion of the shock wave. This should be compared with the case for
$L_{\nu} = 5.5\cdot 10^{52}~\mbox{erg/s}$, which does not produce any explosion. In this case, 
the average shock radius is settled to the value $\sim 10$\% larger than the initial radius after 
$\sim 150$~ms while the net heating rate is fluctuated around a constant value during the same
period.

\begin{figure}
\epsscale{1.0}
\plotone{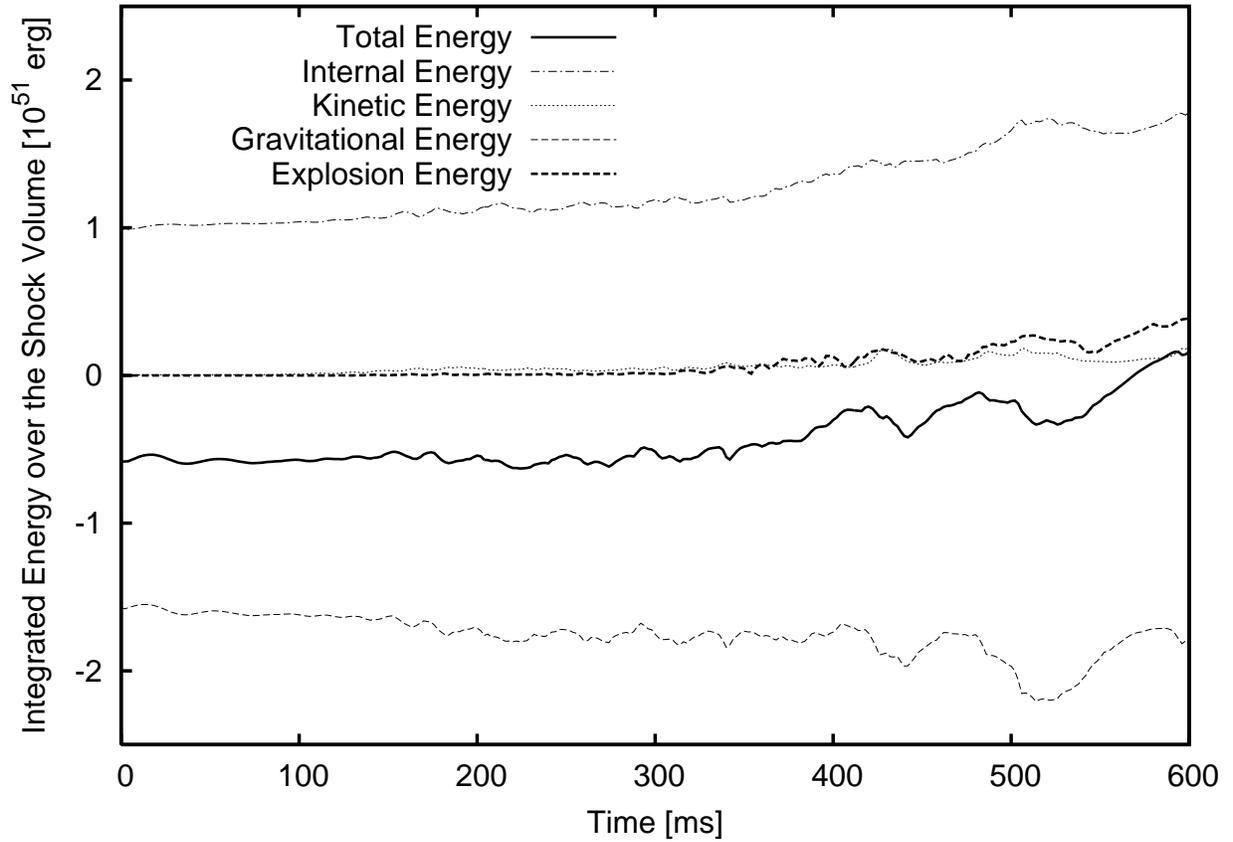}
\caption{Temporal evolutions of various energies integrated over the region inside the shock wave
for the model with $L_{\nu} = 6.0\cdot 10^{52}$ erg/s. 1\% of the $\ell=1$ single-mode velocity 
perturbation is initially added. See the text for the definition of the explosion 
energy.}
\label{fig:energy_Ln60_steady}
\end{figure}

Figure~\ref{fig:energy_Ln60_steady} shows the temporal evolutions
of various energies integrated over the region inside the shock wave 
for the model with $L_{\nu} = 6.0\cdot 10^{52}~\mbox{erg/s}$. 
The explosion energy shown in the figure is defined as the total energy
summed only over the fluid elements which have a positive total energy
and a positive radial velocity simultaneously, as was done, for example, by \citet{yamasawa}.
It is again clearly seen that the increase of the internal energy after $\sim 200$~ms is
responsible for the shock revival. Note that the explosion energy remains zero for a while after the 
shock wave begins its continuous expansion, implying that there is still 
no fluid element with a positive total energy and radial velocity at that time, and that it is 
only after another $\sim 100$~ms that some of the fluid elements gain enough energy to
be ejected. 

\section{Conclusion\label{sec4}}

We have done two-dimensional simulations of SASI in the supernova core
to study its linear growth and nonlinear saturation in detail together with its role in the shock revival.
We employed the realistic EOS and neutrino interaction rates for emission and absorption on nucleons. 
The systematic numerical experiments have been conducted, varying neutrino luminosities from the 
central object and solving spherically symmetric steady accretion flows through the stalled shock wave
as unperturbed initial conditions. We have studied the models both with and without a negative entropy-gradient
in the initial configurations. We have done the mode analysis based on the spherical harmonics decomposition.
Not only the $\ell=1$ single-mode velocity perturbation with an amplitude of 1\% but also the random
multi-mode velocity perturbation with an amplitude $<$~1\% everywhere has been investigated.

We have demonstrated that the non-radial instability grows, indeed,  exponentially in all the models. 
In a model with a relatively high neutrino luminosity, we have found a continuous expansion of the shock wave, 
which will eventually lead to explosion. Since this model did not produce explosion in the spherically symmetric 
simulations, we can infer that the instability tends to lower the critical neutrino luminosity for the shock revival.

In the mode analysis for the models with the initial $\ell =1$ single-mode velocity perturbation, 
we have clearly seen both the linear and nonlinear growths of the instability. We have obtained the growth rates and
characteristic oscillation frequencies by fitting the numerical data and found that the oscillation frequency reflects
the period of the cycle of the inward advections of the velocity- and entropy-fluctuations and the outward
propagations of pressure fluctuations. This fact suggests that the instability is indeed driven by such cycles.
This is, however, in sharp contrast with the recent findings by \citet{blondin_05} that 
the repeated propagations of pressure fluctuations are solely responsible for the instability. 
This apparent discrepancy may be ascribed to the difference of the treatments of neutrino cooling. 

We have observed for the models with the random multi-mode velocity fluctuation that the linear phase 
lasts for $\sim 100$~ms, in which the modes with $\ell=1, 2$ are dominant. It is also interesting that 
in the nonlinear phase, where the nonlinear couplings generate all modes and the saturation occurs, the 
modes with $\ell=1,2$ are still dominant. Detailed analyses of the mode couplings will be necessary to reveal
the physics behind this result. In the models with a negative entropy-gradient, the broadening of the spectra 
has been observed, which we have ascribed to the convective motions. This appears to be supported by 
the recent analysis of the convection in the accretion flows in a supernova core by \citet{foglizzo05}. 
Applying their results to our models, we have shown that the convection plays a minor role in our models. 
The relative significance of SASI and convection, however, depends on the situations and SASI appears to be dominant 
for relatively low neutrino luminosities \citep{scheck_04}. It is important to note here that the growth of SASI
is rather slow and takes $\sim 100$~ms. Although we have found a shock revival due to the SASI-enhanced
neutrino heatings in some cases as mentioned above, whether we find such a revival in the reality crucially 
depends on if the neutrino luminosity can be sustained for this period, not an easy job.

If SASI alone cannot get the job done, we will have to find something more to boost the shock revival. 
We are currently working on some combinations of microphysics and SASI and will soon publish the results
elsewhere \citep{ohnishi05}. The influence of rotation on SASI should be addressed also and 
may be important for the possible correlation of the kick velocity and
the rotation axis \citep{Lai,ohnishi05}.

\acknowledgements{We are grateful to K. Sumiyoshi for valuable
discussions and for providing us
with the tabulated equation of state, which can be handled
without difficulties.
We also thank
T. Yamasaki and M. Watanabe for providing us
with the information about their models.
K.K. expresses thanks to K. Sato for continuing encouragements. 
The numerical calculations were partially done on the
supercomputers in RIKEN and KEK (KEK supercomputer Projects No.02-87 and
No.03-92). This work was supported in part by the Japan Society for
Promotion of Science(JSPS) Research Fellowships (K.K.), 
Grants-in-Aid for the Scientific Research from the Ministry of
Education, Science and Culture of Japan (No.S14102004, No.14079202, 
No.17540267), and Grant-in-Aid for the 21st century COE program
``Holistic Research and Education Center for Physics of Self-organizing Systems''.}

\clearpage

\clearpage

\clearpage

\clearpage

\clearpage

\clearpage

\clearpage

\clearpage

\clearpage

\clearpage

\begin{deluxetable}{cccccccc}
\tabletypesize{\scriptsize}
\tablecaption{Key Variables in SASI.\label{table:growth_rate}}
\tablewidth{0pt}
\tablehead{
\colhead{$L_{\nu}$ ($10^{52}$ erg/s)}
 & \colhead{$\gamma$ (s$^{-1}$)}
 & \colhead{$\omega$ (s$^{-1}$)}
 & \colhead{$R_{\rm s,equil}$ ($10^{5}$ cm)}
 & \colhead{$w_{\rm s}$ ($10^{5}$ cm)}
 & \colhead{$\omega_{\rm adv}$ (s$^{-1}$)}
 & \colhead{$\omega_{\rm snd}$ (s$^{-1}$)}
 & \colhead{$\omega_{\rm cyc}$ (s$^{-1}$)}}
\startdata
 3.0 & 42.4 & 915 & 66.9 & 30.9 & 1098 & 5867 & 925\\
 5.5 & 45.7 & 277 & 128 & 79.2 & 262 & 1832 & 229\\
 6.0 & 38.3 & 188 & 144 & 93.5 & 207 & 1497 & 181\\
 6.5 & 35.6 & 143 & 167 & 114 & 159 & 1199 & 141\\
\enddata
\tablecomments{%
$L_{\nu}$ represents the model luminosities.
The growth rate and the oscillation frequency denoted as  $\gamma$ and $\omega$, 
respectively, are obtained by the least square fitting to
the numerical results in the linear regime.
$R_{\rm s,equil}$ is the initial shock radius
and $w_{\rm s}$ is the distance between the shock radius
and the neutrino sphere; $w_{\rm s} = R_{\rm s,equil} - r_{\nu}$.
The frequencies associated with the advection and the sound-propagation between the 
shock and the neutrino sphere are denoted as $\omega_{\rm adv}$ and $\omega_{\rm snd}$, 
respectively, and are defined as
$\omega_{\rm adv} = 2\pi/\int_{r_{\nu}}^{R_{\rm s}}(1/v_r)dr$ and
$\omega_{\rm snd} = 2\pi/\int_{r_{\nu}}^{R_{\rm s}}(1/c_{\rm s})dr$,
respectively. They are evaluated numerically for the initial conditions.
The characteristic frequency of SASI is given by the cycle frequency, 
$\omega_{\rm cyc} = 2\pi/
[\int_{r_{\nu}}^{R_{\rm s}}(1/v_r)dr+
 \int_{r_{\nu}}^{R_{\rm s}}(1/c_{\rm s})dr$]. See the text for more detail.}
\end{deluxetable}

\end{document}